\documentclass[onecolumn,showpacs,preprintnumbers,amsmath,amssymb]{revtex4}
\usepackage{graphicx}
\usepackage{dcolumn}
\usepackage{bm}
\usepackage{epsfig}
\usepackage{subfigure}

\newcommand{\bi}{\begin{itemize}}
\newcommand{\ei}{\end{itemize}}
\newcommand{\be}{\begin{eqnarray}}
\newcommand{\ee}{\end{eqnarray}}
\newcommand{\beq}{\begin{equation}}
\newcommand{\eeq}{\end{equation}}
\newcommand{\beqn}{\begin{equation*}}
\newcommand{\eeqn}{\end{equation*}}
\newcommand{\bbmatrix}{\left( \begin{array}}
\newcommand{\eematrix}{\end{array} \right)}

\begin{document}
\title{The optimal Boson energy for superconductivity in the Holstein model}
\author{Chungwei Lin\footnote{email: clin@merl.com}, Bingnan Wang, and Koon Hoo Teo}
\affiliation{Mitsubishi Electric Research Laboratories, 201 Broadway, Cambridge, MA 02139, USA} 
\date{\today}

\begin{abstract}
We examine the superconducting solution in the Holstein model, where 
the conduction electrons couple to the dispersionless Boson fields, using 
the Migdal-Eliashberg  theory and Dynamical Mean Field Theory.
Although different in numerical values, both methods imply the existence of an optimal Boson energy
for superconductivity at a given electron-Boson coupling. 
This non-monotonous behavior can be understood as an interplay between the polaron and superconducting physics,
as the electron-Boson coupling is the origin of the superconductor, but at the same time traps the conduction
electrons making the system more insulating.
Our calculation provides a simple explanation on the recent experiment on sulfur hydride, 
where an optimal pressure for the superconductivity was observed. 
The validities of both methods are discussed. 
\end{abstract}

\maketitle

\section{Introduction}

Since the discovery of superconductivity by Onnes \cite{Onnes}, 
countless efforts have been dedicated to understanding the microscopic origin 
of the phenomena, as well as to searching/synthesizing materials of high superconducting critical temperatures ($T_c$).
Based on the microscopic theories, the superconductors are classified as 
``conventional'' and ``unconventional'' superconductors. 
The former class can be well described by the 
BCS (Bardeen-Cooper-Schrieffer) theory \cite{PhysRev.104.1189, PhysRev.106.162, PhysRev.108.1175} 
or its variants \cite{Tc_eps}, whereas the latter class is still controversial in its microscopic mechanisms \cite{HighTc_In_Suspence}.
The unconventional superconductors include the layered materials such as cuprates 
\cite{Cuprate_Bednorz_Muller, RevModPhys.75.473} and iron-based materials \cite{doi:10.1021/ja063355c, doi:10.1021/ja800073m, FeAs_Film_100K_STO_2015}.
The multi-orbital nature and strong electron correlation intrinsically complicate the problem, 
as there can be several competing phases \cite{RevModPhys.66.763, RevModPhys.70.1039}.
For this reason, searches of unconventional high temperature superconductors are mainly based on 
``perturbing the existing superconductors'' (via doping, applying pressure, interfacing with other materials ... etc) 
and ``exhausting all possible compounds''  \cite{1468-6996-16-3-033503}.
The searches of conventional high temperature superconductors, on the other hand, 
are essentially guided by the BCS theory, or by the more realistic Eliashberg model 
\cite{Eliashberg_1960, Migdal_1958, PhysRev.125.1263, PhysRevB.12.905, Z.Phys.263.59, Grimvall, Allen_SC},
which is different from BCS theory in its explicit inclusion of the phonon (or the Boson in general) degrees  of freedom.
In the Eliashberg model, the origin of the effective electron-electron attraction is the electron-phonon coupling, 
and the main factor against superconductivity, the Coulomb repulsion,
is treated at a semi-empirical level by one parameter \cite{PhysRev.125.1263, PhysRev.167.331, Allen_SC}.
Therefore, the key to enhance $T_c$ is to control the phonon-related parameters -- the Debye frequency and the electron-phonon coupling. 
Recent experiments on sulfur hydride, whose motivation behind is to increase
the Debye frequency by using light elements (H) and applying high pressure, exhibit 
a record superconducting $T_c$ at 203 K \cite{Nature525.73.2015}. 
The strong enhancement of superconducting $T_c$ in the mono-layer FeAs or FeSe on SrTiO$_3$ substrate
is deeply related to the interfacial optical phonon mode 
\cite{0256-307X-29-3-037402, FeAs_Film_100K_STO_2015, FeAs_Film_STO_2014, 1367-2630-18-2-022001, 0953-2048-29-5-054009}.
Attempts of using  Boson  other than phonons, such as the plasma in meta-materials, to mediate 
the electron-electron attraction also appear promising \cite{SciRep_2014, PhysRevB.91.094501}.

The Holstein model \cite{Holstein_1959}, where the conduction electrons couple to the dispersionless Boson fields,
is the simplest model that captures the physics of conventional superconductors. 
In this work, we examine the superconducting solution in Holstein model, using the 
Migdal-Eliashberg (ME)  theory  \cite{Migdal_1958, Eliashberg_1960} and the 
Dynamical Mean Field Theory (DMFT) \cite{RevModPhys.68.13, Kotliar_04, RevModPhys.77.1027, RevModPhys.83.349}
with the exact diagonalization (ED) impurity solver.
This model has been intensely studied 
\cite{PhysRevB.48.6302, PhysRevB.50.6939, PhysRevLett.75.2570, PhysRevB.58.14320, PhysRevB.65.224301, PhysRevB.56.4513, PhysRevB.58.3094, PhysRevLett.89.196401, PhysRevLett.99.146404, 
PhysRevLett.109.176402, PhysRevB.91.045128, PhysRevB.46.271, 0295-5075-85-5-57003}, but relatively few 
explicitly break the gauge symmetry to obtain the superconducting solution \cite{0953-8984-17-37-005,PhysRevLett.113.266404}.
Although different in numerical values, both ME and DMFT imply the existence of an optimal Boson energy
for superconductivity at a given electron-Boson coupling.  
The existence of the optimal Boson energy can be expected from the BCS theory -- 
if we take the cutoff energy as the Boson energy $\Omega$, and the effective electron-electron attraction as 
$-g^2/\Omega$ (an estimate from the second-order perturbation) with $g$ being the electron-Boson coupling, the superconducting gap 
behaves as $\Delta (\Omega) \sim \Omega \exp [- (g^2/\Omega) D_0 ] $ with $D_0$ the electron density of states at the Fermi energy. 
The ME theory actually gives a very similar behavior. The DMFT, however, gives different ground state under some parameter regimes,
as it captures the polaron effect, where the Boson that mediates the electron-electron attraction can make the system insulating.  
The DMFT calculation also elucidates the relationship between the polaron solution and the superconducting solution.

The rest of the paper is organized as follows.
In Section II we describe the Holstein model and the two methods -- the ME theory 
and DMFT -- to obtain the superconducting solutions.
We provide a simple picture that is emerged from DMFT, on how the Boson field can lead to the superconducting solution.
In Section III, we present our main results, compare them to those in the literature, and discuss their validities and implications.
Finally a brief conclusion is given in Section IV.

\begin{figure}[htp]
\epsfig{file=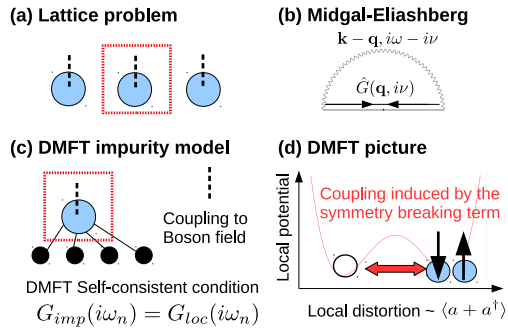, width = 0.4\textwidth}
\caption{
(a) The Holstein model in real space: each local electron couples to  an external Boson field.
Each dashed line represents the electron-Boson coupling.
(b) The only Feynman diagram included in the Migdal-Eliashberg theory, with the 
lattice Green's function $\hat{G}(\mathbf{k}, i\omega_n)$ solved self-consistently. 
(c) The auxiliary impurity model in DMFT. Instead of solving the lattice problem 
where each local orbital couples to an external Boson field,
DMFT iteratively solves an impurity problem where only the impurity orbital couples to 
the external Boson field. (d) The picture emerged from the DMFT calculation. 
If the doubly-occupied and zero-occupied impurity orbitals are degenerate or close in energy 
for the local Hamiltonian, introducing a coupling between these two local states by breaking the particle conservation further
lowers the energy via producing a ``binding'' combination of these two states. 
}
\label{fig:Model-method-picture}
 \end{figure}

\section{Holstein model and solvers} 
In this section we introduce the Holstein model, and the two solvers we used 
-- the ME theory and DMFT -- to solve this model. 
The expressions of the main observables, including the superconducting gap 
and the pairing amplitude, are given.
Several hints of the existence of the optimal Boson energy for superconductivity 
will be highlighted.
 
\subsection{Holstein model and superconducting gap}

The Holstein model is given by
\beq 
\begin{split}
H &= H_{elec} + H_{ph} + H_{e-ph}  \\
&=
\sum_{\mathbf{k}} \varepsilon_{\mathbf{k}} (c^\dagger_{\mathbf{k}, \uparrow} c_{\mathbf{k}, \uparrow} 
+ c^\dagger_{\mathbf{k}, \downarrow} c_{\mathbf{k}, \downarrow}) +  \Omega \sum_{\mathbf{k}} a^\dagger_{\mathbf{k}} a_{\mathbf{k}}
+ \frac{g}{\sqrt{N}} \sum_{\mathbf{k}, \mathbf{q}} 
(c^\dagger_{\mathbf{k}+\mathbf{q}, \uparrow} c_{\mathbf{k}, \uparrow} 
+ c^\dagger_{\mathbf{k}+\mathbf{q}, \downarrow} c_{\mathbf{k}, \downarrow}) (a_{\mathbf{q}} + a^\dagger_{-\mathbf{q}}).
\end{split}
\label{eqn:Holstein}
\eeq 
Here $c_{\mathbf{k}, \sigma} $ represents the Fermion degree of freedom, whereas 
$a_{\mathbf{k} }$ represents the Boson degree of freedom.
In Eq.~\eqref{eqn:Holstein}, we can rewrite
$H_{ph}$ and $H_{e-ph}$ in the real-space coordinate as 
\beq 
H_{ph} + H_{e-ph} =  \Omega \sum_{i} a^\dagger_{i} a_{i}
+ g \sum_{i, \sigma} (c^\dagger_{i, \sigma} c_{i, \sigma} - 1)  (a_{i} + a^\dagger_{i}).
\label{eqn:Holstein_real_0}
\eeq 
This form is more natural for DMFT calculations. 
The Holstein model in the real-space representation is illustrated in Fig.~\ref{fig:Model-method-picture}(a).
In this work, we shall consider the conduction band of semicircular density of states (DOS)
$\nu(\varepsilon) = \sqrt{4 t^2 - \varepsilon^2}/(2 \pi t^2)$. This corresponds to 
the Bethe lattice of infinite dimension, a limit where the DMFT result becomes exact.
The bandwidth is fixed at $4t$ with $t=1$, and all energy scales, including the electron-Boson coupling $g$
and the Boson energy $\Omega$, are measured in $t$. 
We only show  the results at half filling, and the -1 in 
$(c^\dagger_{i, \sigma} c_{i, \sigma} - 1)$ of Eq.~\eqref{eqn:Holstein_real_0} ensures that  
the Boson field is at its ground state when the local occupation is 1 (half-filled) \cite{PhysRevLett.99.146404}.

To obtain the superconducting solutions, both ME theory and DMFT 
self-consistently determine the Nambu Green's functions at the Matsubara frequencies. 
Defining a Nambu spinor as $\Psi^{\dagger}_{\mathbf{k}} = (c^{\dagger}_{\mathbf{k}, \uparrow}, c_{-\mathbf{k}, \downarrow})$, 
the lattice Green function (a 2$\times$2 matrix) on the imaginary-time axis and at the Matsubara frequency is given by
\beq
\begin{split}
\hat{G} (\mathbf{k}, \tau) &= -T \langle  \Psi_{\mathbf{k}}(\tau) \Psi^{\dagger}_{\mathbf{k}}(0) \rangle =
\begin{pmatrix} G(\mathbf{k}, \tau) & F(\mathbf{k}, \tau) \\ F(\mathbf{k}, \tau)^* & -G(-\mathbf{k},-\tau) \end{pmatrix} \\
\Rightarrow &  
\hat{G}(\mathbf{k}, i \omega_n) = 
\begin{pmatrix} G(\mathbf{k},  i \omega_n) & F(\mathbf{k},  i \omega_n) \\ 
F^*(\mathbf{k}, i \omega_n) & -G(-\mathbf{k},- i \omega_n) \end{pmatrix}
  \end{split}
  \label{eqn:Nambu_SC}
\eeq
where $\langle ... \rangle$ represents the ground state expectation value. Using Pauli matrices 
\beq 
\sigma_1 = \begin{pmatrix} 0 & 1 \\ 1 & 0 \end{pmatrix},\,\,
\sigma_2 = \begin{pmatrix} 0 & -i \\ i & 0 \end{pmatrix},\,\,
\sigma_3 = \begin{pmatrix} 1 & 0 \\ 0 & -1 \end{pmatrix},
\eeq 
the self-energies and the lattice Green's functions are parameterized as
\beq 
\begin{split}
\hat{\Sigma} (\mathbf{k}, i \omega_n) &= 
i \omega_n [1- Z(\mathbf{k}, i \omega_n)] \hat{\sigma}_0 +  \chi (\mathbf{k}, i \omega_n) \hat{\sigma}_3
+ \phi (\mathbf{k}, i \omega_n) \hat{\sigma}_1 + \phi_2 (\mathbf{k}, i \omega_n) \hat{\sigma}_2 \\
\hat{G}^{-1} (\mathbf{k}, i \omega_n) &= 
\hat{G}^{-1}_0 (\mathbf{k}, i \omega_n)  - \hat{\Sigma}  (\mathbf{k}, i \omega_n) \\
&= i \omega_n Z(\mathbf{k}, i \omega_n) \hat{\sigma}_0 - [\varepsilon_{\mathbf{k}} + \chi (\mathbf{k}, i \omega_n)] \hat{\sigma}_3
- \phi (\mathbf{k}, i \omega_n) \hat{\sigma}_1 - \phi_2 (\mathbf{k}, i \omega_n) \hat{\sigma}_2. 
\end{split}
\label{eqn:G_Inverse_full}
\eeq 
with $\hat{G}_0 (\mathbf{k}, i \omega_n)$ being the non-interacting Green's function. 
Without loss of generality, we can choose $\phi_2 = 0$, and the task is to determine 
$Z(\mathbf{k}, i \omega_n)$, $\phi (\mathbf{k}, i \omega_n)$, and $\chi (\mathbf{k}, i \omega_n)$ 
self-consistently using some approximation. 

When the lattice Green's function are obtained, its poles determine the single-particle excitations. 
From Eq.~\eqref{eqn:G_Inverse_full}, the poles are determined by (with the analytical continuation $i \omega_n \rightarrow \omega$)
\be
-[\omega Z(\mathbf{k}, \omega)]^2 + [\varepsilon_{\mathbf{k}} + \chi (\mathbf{k},  \omega)]^2
+ \phi^2 (\mathbf{k}, \omega)  &=& 0.
\ee 
For  the normal state ($\phi = 0$), the poles are given by 
$\omega = \pm [\varepsilon_{\mathbf{k}} + \chi (\mathbf{k},  \omega) ]/Z(\mathbf{k}, \omega)$, which 
are simply the quasi-particle (quasi-hole) excitations. 
For the non-zero $\phi$, the excitation occurs at 
$\omega = \pm \sqrt{ [\varepsilon_{\mathbf{k}} + \chi (\mathbf{k},  \omega) ]^2 + \phi (\mathbf{k},  \omega)^2  } /Z(\mathbf{k}, \omega)$.
When neglecting the $\mathbf{k}$ dependence (an approximation we are using in this paper), the superconducting gap is 
obtained by energy difference $|\omega_+ - \omega_-|$, with $\omega_{\pm} = \pm \phi(\omega) /Z(\omega)$.  
Keeping only the constant term  of $Z$ and $\phi$, the gap is approximately 
\beq 
\frac{\Delta}{2} = \frac{ \phi }{Z} (\omega = 0) \approx \frac{ \phi (i \omega_0)}{Z(i \omega_0)},
\label{eqn:gap}
\eeq 
where $Z(0) \approx Z(i \omega_0)$ and $\phi(0) \approx \phi(i \omega_0)$ are used.
In addition to the superconducting gap, the superconductivity can also be characterized by 
the pairing amplitude $\Psi$
\beq 
\Psi \equiv \langle c_{i,\uparrow} c_{i,\downarrow}  \rangle = T \sum_n \phi (i \omega_n).
\label{eqn:off-diag}
\eeq 
Note that $\Psi$ is a dimensionless quantity, whose amplitude is always smaller than one.
Eq.~\eqref{eqn:gap} and Eq.~\eqref{eqn:off-diag} will be used to characterize the superconducting state.

\subsection{Migdal-Eliashberg theory}
The Migdal-Eliashberg theory is formulated in the momentum space. 
It keeps only the ``Fock'' contribution in the self energies [see Fig.~\ref{fig:Model-method-picture}(b) 
for the diagrammatic representation]:  
\beq 
\begin{split}
 \hat{\Sigma}  (\mathbf{k}, i \omega_n) 
   &=- T \sum_{\mathbf{k}', n'} \hat{\sigma}_3 \hat{G} (\mathbf{k}', i \omega_n) \hat{\sigma}_3 
  \times   |g(\mathbf{k}, \mathbf{k}')|^2 D (\mathbf{k}- \mathbf{k}', i \omega_n - i \omega_{n'}),  
 \end{split}
 \label{eqn:Sigma_0}
\eeq
$\hat{G}^{od}$ is the off-diagonal part of the Green's function, 
$g(\mathbf{k}, \mathbf{k}')$ is the coupling that annihilates an electron of momentum $\mathbf{k}$ and creates
an electron of momentum $\mathbf{k}'$,
and $D(\mathbf{k}, i \omega_n)$ is the Boson Green's function.
Substituting Eq.~\eqref{eqn:Sigma_0} into Eq.~\eqref{eqn:G_Inverse_full}, we obtain the equation for 
$Z(\mathbf{k}, i \omega_n)$, $\phi(\mathbf{k}, i \omega_n)$, and $\chi(\mathbf{k}, i \omega_n)$ as 
\begin{subequations}
\begin{align}
[ 1 - Z(\mathbf{k}, i \omega_n) ] i \omega_n &= -T \sum_{\mathbf{k}', n', \lambda} |g(\mathbf{k}, \mathbf{k}', \lambda)|^2 
D_\lambda (\mathbf{k}- \mathbf{k}', i \omega_n - i \omega_{n'}) 
\frac{ i \omega_{n'} Z(\mathbf{k}', i \omega_{n'})  }
{-\Theta(\mathbf{k}', i \omega_{n'})},  \\ 
\chi (\mathbf{k}, i \omega_n) &= -T \sum_{\mathbf{k}', n', \lambda} |g(\mathbf{k}, \mathbf{k}', \lambda)|^2 
D_\lambda (\mathbf{k}- \mathbf{k}', i \omega_n - i \omega_{n'})
\frac{ [\varepsilon_{\mathbf{k}'} + \chi (\mathbf{k}', i \omega_{n'})]   }
{-\Theta(\mathbf{k}', i \omega_{n'})},   \\
\phi (\mathbf{k}, i \omega_n) &= T \sum_{\mathbf{k}', n', \lambda} |g(\mathbf{k}, \mathbf{k}', \lambda)|^2 
D_\lambda (\mathbf{k}- \mathbf{k}', i \omega_n - i \omega_{n'})
\frac{ \phi  (\mathbf{k}', i \omega_{n'})   }
{-\Theta(\mathbf{k}', i \omega_{n'})},
\end{align}
\label{eqn:Eliashberg_general}
\end{subequations}
where $\Theta(\mathbf{k},i \omega_n) = [\omega_n Z(\mathbf{k},i \omega_n) ]^2 + \varepsilon (\mathbf{k})^2
+ \phi(\mathbf{k},i \omega_n)^2 + \chi (\mathbf{k},i \omega_n)^2$.
For the Holstein model defined in Eq.~\eqref{eqn:Holstein}, we have 
$g(\mathbf{k}, \mathbf{k}') = g$ and $D (\mathbf{q}, \omega  ) = -\frac{2 \Omega}{\omega^2 + \Omega^2  }$. 
We further simplify the equation by neglecting the momentum dependence, i.e.
$Z(\mathbf{k}, i \omega_n) \rightarrow Z(i \omega_n) \equiv Z_n$, 
$\phi(\mathbf{k}, i \omega_n) \rightarrow \phi(i \omega_n) \equiv \phi_n$
$\chi(\mathbf{k}, i \omega_n) \rightarrow \chi(i \omega_n) \equiv \chi_n$, and obtain the coupled equations  
\be 
[\omega_n Z_{n}] &=& \sum_{n'} -K(n,n')  [\omega_{n'} Z_{n'}] + \omega_n \nonumber \\
\chi_n &=& +K(n,n') \chi_{n'} + C_n
\nonumber \\ 
\phi_n &=& \sum_{n'} -K(n,n') \phi_{n'}
\label{eqn:simplified_eq}
\ee 
with  $K(n,n') = -T g^2 \frac{2 \Omega}{\Omega^2+( \omega_n-\omega_{n'} )^2} \times  \sum_{\mathbf{k} } 
\frac{1/N}{\Theta(\mathbf{k},i \omega_{n'} )}$, 
$C_n = +T g^2  \frac{2 \Omega}{\Omega^2+( \omega_n-\omega_{n'} )^2} \times \sum_{\mathbf{k} }
\frac{\varepsilon (\mathbf{k})/N }{\Theta(\mathbf{k},i \omega_{n'} )}$, and  
$\Theta(\mathbf{k},i \omega_n) = (\omega_n Z_n)^2 + \varepsilon^2 (\mathbf{k}) + \phi_n^2 + \chi_n^2$.
With the semicircular DOS $\nu(\varepsilon) = \sqrt{4 t^2 - \varepsilon^2}/(2 \pi t^2)$, 
$K(n,n')$ and $C_n$ are evaluated by 
\be 
K(n,n') &=&  -T g^2 \frac{2 \Omega}{\Omega^2+( \omega_n-\omega_{n'} )^2}  \times \int d\varepsilon 
\frac{\nu(\varepsilon)}{\Theta(\varepsilon,i \omega_{n'} )}, \nonumber \\ 
C_n &=& +T g^2  \frac{2 \Omega}{\Omega^2+( \omega_n-\omega_{n'} )^2} \times 
\int d\varepsilon \nu(\varepsilon) \frac{\varepsilon - \mu}{\Theta(\varepsilon,i \omega_{n'} )}, 
\label{eqn:K_C_evulation}
\ee 
with $\Theta(\varepsilon,i \omega_{n} ) = (\omega_n Z_n)^2 + (\varepsilon-\mu)^2  + \phi_n^2 + \chi_n^2$.
We solve Eq.~\eqref{eqn:simplified_eq} by iteration.
The zero-temperature results obtained by using $T=0.001$ and keeping 4000 Matsubara frequencies; 
the results are checked against those obtained using lower temperature and keeping more Matsubara frequencies.
Three comments about Eq.~\eqref{eqn:simplified_eq} are worth noting.
First, by linearizing Eq.~\eqref{eqn:simplified_eq} ($\phi_n$ components), 
det$[K(n,n')+I] = 0$ determines the critical temperature $T_c$.
Second, due to the neglect of momentum dependence in the self energy, 
we expect the superconducting gap obtained using the approximation is under-estimated. 
Finally, as the magnitudes of $K(n,n')$ are small at both small and large $\Omega$,
Eq.~\eqref{eqn:simplified_eq} suggests a optimal $\Omega$ for the superconductivity.
At this stage it is simply a mathematical observation, and we shall give a more physical discussion
on Section II.D and Section III.A.

\subsection{Dynamical mean field theory}
 
Dynamical mean field theory \cite{RevModPhys.78.865, Kotliar_04} fully captures the 
local interaction via an auxiliary impurity model,  and determines the impurity-bath
hybridization parameters by equating the lattice local Green's function to the impurity Green's function
[see Fig.\ref{fig:Model-method-picture}(c) for an illustration]. 
For the superconducting solution, the impurity model is 
\beq 
\begin{split}
H_{imp,SC} &= \varepsilon_d  \sum_{\sigma} c^{\dagger}_{1,\sigma} c_{1,\sigma}
+ \sum_{p=1}^N t_{sc,p} (c^{\dagger}_{p,\uparrow} c^{\dagger}_{p,\downarrow} + h.c.)
+ \sum_{p=2, \sigma}^{N}  t_p [ c^{\dagger}_{1,\sigma} c_{p,\sigma} + h.c.]
+ \sum_{p=2, \sigma}^{N}  \varepsilon_p  c^{\dagger}_{p,\sigma} c_{p,\sigma} \\
& + g ( n_{1, \uparrow} + n_{1, \downarrow} -\alpha ) (a+a^{\dagger} ) + \Omega a^{\dagger}a,
\end{split}
\label{eqn:Imp_Ph}
\eeq 
which explicitly breaks the particle conservation via the term $t_{sc,p} (c^{\dagger}_{p,\uparrow} c^{\dagger}_{p,\downarrow} + h.c.)$.
We have assumed site 1 to be the impurity site. We use exact diagonalization (ED) \cite{Liebsch_12} for the impurity problem,
and consider the zero-temperature solution.
Due to computational cost, we include five bath orbitals (totally six orbitals including the impurity), which 
is shown to be sufficient for the attractive Hubbard model \cite{PhysRevB.88.035123}. 
As the particle number is not conserved, the impurity problem is solved in the grand-canonical ensemble.
The details can be found in Ref.~\cite{PhysRevB.88.035123}, and we point out one aspect specific to the Boson degree of freedom.
As the model includes both Fermions and Bosons, the Hilbert space of Eq.~\eqref{eqn:Imp_Ph} is defined as $| m \rangle_e \otimes | n \rangle_{ph}$. 
The electronic state $| m \rangle_e$ is a Fock state built from creating the Bogoliubov particles 
on the $| 0 \rangle \equiv \Pi_p c^{\dagger}_{p,\downarrow} | \mbox{vac} \rangle$, i.e.
$ | m \rangle_e = \Pi_i \gamma^\dagger_i |0\rangle $ (with Bogoliubov orbitals $\gamma^\dagger_i$ being composed of 
$c^{\dagger}_{\uparrow}$ and $c_{\downarrow}$),  
whereas the phonon state is built from 
$| n \rangle_{ph} \sim (a^\dagger)^n | 0 \rangle_{ph}$. In principle, there are infinite 
number of phonon states; in practice we keep $n_{max}$ phonon states ($| n \rangle_{ph}$ with $n=0$ to $n_{max}-1$) 
such that the converged result does not change for $n_{max} \rightarrow n_{max}+5$.
A simple criterion is that $n_{max} \Omega$ is much larger than all relevant energy scales such as 
bandwidth and electron-Boson coupling, therefore 
the smaller the Boson energy $\Omega$ is, the more phonon states one needs to keep.
For this reason the parameter space of small $\Omega$ is difficult to reach. 
Typically we use $n_{max}$ ranging from  20 to 40. 
Two technical details are also noted.
First, in the calculation, an effective temperature is needed, and we choose $T_{eff} =0.01$, 
based on which 1000 Matsubara frequencies are kept. 
Second, we use configuration interaction impurity solver \cite{PhysRevB.86.165128, PhysRevB.88.035123, PhysRevB.90.235122, PhysRevLett.114.016402,
PhysRevB.92.155135}  in the early self-consistency iterations, which significantly accelerates the convergence.


We complete the discussion by examining how the coupling to a  Boson field can lead to the superconducting 
solution in DMFT, as the electron-electron attraction may not be obvious in the model.
In the impurity model [Eq.~\eqref{eqn:Imp_Ph}], double and zero occupation on the impurity orbital result in 
phonon Hamiltonians of $\pm g(a+a^\dagger) + \Omega a^\dagger a 
= -\frac{g^2}{\Omega} + \Omega (a^\dagger \pm g/\Omega)( a\pm g\Omega)$. 
Regardless of the sign of $g$, they both gain energy of $-g^2/\Omega$.
If we ignore the Boson dynamics and use the semiclassical impurity solver 
\cite{PhysRevB.54.5389,PhysRevB.71.235113, PhysRevB.79.205109}, 
the system is trapped in one of the minimum and we obtain a polaron solution --
half of the lattice sites are doubly occupied and half the them empty.
The superconducting state, which allows a direct coupling between these two local minimum
whose local occupations are differed by two, 
further lowers the energy via producing a ``binding'' combination of these two states. 
This picture, emerged from the DMFT formalism that emphasizes the local physics, 
connects the polaron solution and superconducting solution,
and is illustrated in Fig.~\ref{fig:Model-method-picture}(d).

\subsection{Polaron and superconductivity}

\begin{figure}[htp]
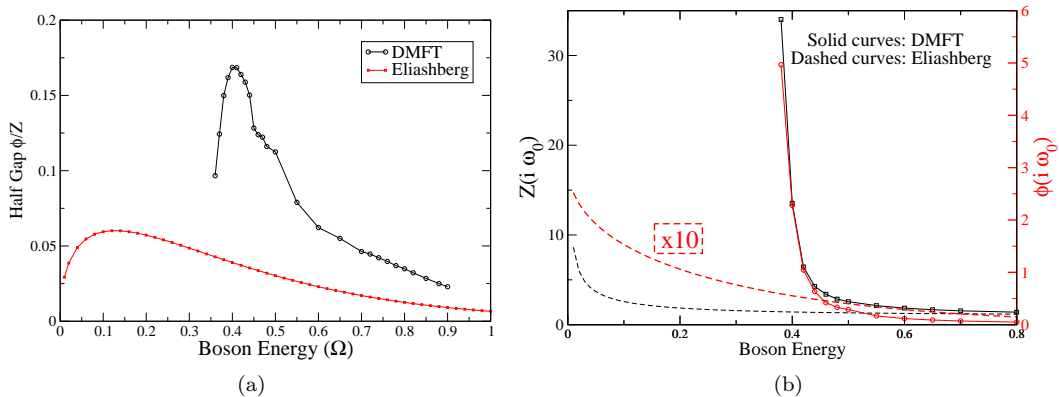

\subfigure[]{ \epsfig{file=Order_t1.0_U0.00_gg0.60_ww0.3-0.9_T0.010_Mw800_Ch0.50_Ne6_Ns6_Full.eps, width = 0.36\textwidth} }
\subfigure[]{ \epsfig{file=ZPhi_Observable_g0.6_ch0.5_both.eps, width = 0.4\textwidth} }
\caption{(a) The half gap, computed from both DMFT and Migdal-Eliashberg theory,
as a function of Boson energy for the Holstein model at half filling. 
The electron-boson coupling $g = 0.6$. 
For both Migdal-Eliashberg theory and the DMFT, there exists an optimal Boson energy $\Omega_{opt}$
for the superconducting gap: $\Omega_{opt} \sim 0.4$ for DMFT; $\Omega_{opt} \sim 0.15$ for the Eliashberg theory.
The DMFT predicts a larger gap than the Migdal-Eliashberg theory at large $\Omega$.
(b) $Z(i \omega_0)$ (black, left y-axis) and $\phi(i \omega_0)$ (red, right y-axis) as 
a function of Boson energies, obtained from DMFT (solid curves) 
and the ME theory (dashed). $\phi(i \omega_0)$ of ME theory 
are multiplied by 10 to fit the scale.
When $\Omega$ decreases, both $Z(i \omega_0)$  and $\phi(i \omega_0)$ increase.
The ME theory gives a milder behavior compared to DMFT.
}
\label{fig:Half_Gap_g0.6}
 \end{figure}

We conclude this section by associating the polaron and superconducting effects to different components of the Green's function,
based on which these two effects can be quantified.
In Eq.~\eqref{eqn:G_Inverse_full}, the superconductivity is characterized by the off-diagonal component $\phi$, whereas 
the polaron effect by the diagonal component $Z$. 
The superconducting part is straightforward, as $\phi$ directly relates to the pairing amplitude.  
For the polaron part, we first note that $1/Z$ is the quasi-particle weight, which is always between 0 and 1 \cite{Mahan}.
Smaller $1/Z$ leads to a smaller spectral weight near the Fermi energy ($E_F$, which is zero in our convention).
The coupling to Bosons tends to slow the electron motion (regardless of its spin), as the excited Bosons  ``drag'' the electron motion. 
This polaron effect makes the system more insulating, and results in the large $Z$ and the reduced spectral function. 
The polaron effect is expected to become stronger when the Boson is easier to excite, which happens at smaller Boson energies $\Omega$.
Our simulations (both ME theory and DMFT) indeed give a larger $Z$ in these regions [see Fig.~\ref{fig:Half_Gap_g0.6}(b)].
As the reduction in DOS is against superconductivity, a reduction of superconductivity at smaller $\Omega$ 
is expected, and is indeed observed for both methods [see Section III.A].

\section{Results and discussion}

 In this section we show the numerical results and discuss their implications.
For both Migdal-Eliashberg theory and DMFT, the superconducting gap [$\phi/Z$, Eq.~\eqref{eqn:gap}] and the pairing amplitude
[$\Psi$, Eq.~\eqref{eqn:off-diag}] are shown.  
For DMFT, the computed spectral functions are also shown.

\begin{figure}[htp]
\vspace{0.05\textwidth}
\epsfig{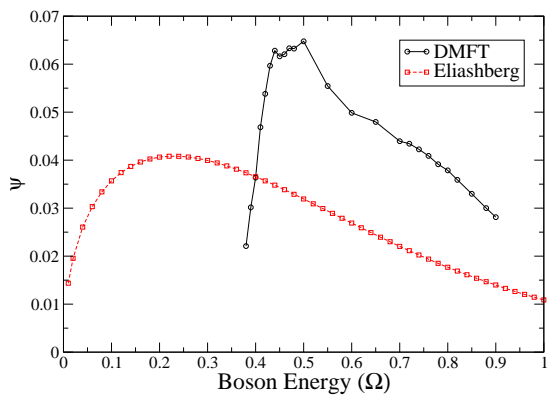} 
\caption{ The pairing amplitude, $\Psi \equiv \langle d_\uparrow d_\downarrow \rangle$, as a function of the Boson energy $\Omega$.
Results obtained from both DMFT and Eliashberg theory are shown.
The shape similar the superconducting gap [Fig.~\ref{fig:Half_Gap_g0.6}(a)] is observed.
}
\label{fig:Ordere_g0.6}
 \end{figure}

\subsection{Superconducting gap, pairing amplitude, and spectral functions}

\begin{figure}[htp]
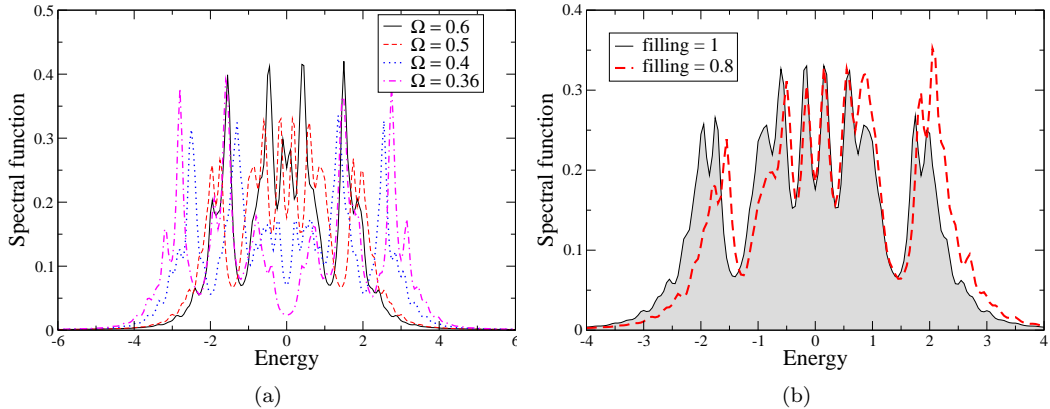

\subfigure[]{\epsfig{file=Spec10I3_t1.0_U0.00_gg0.60_ww0.6-0.4_T0.010_Mw800_Ch0.50_err80.00E-4_dG8.00E-2_Ne6_Ns6_CI6_CASOrb1-12.eps, width = 0.38\textwidth} }
\subfigure[]{\epsfig{file=Spec10I3_t1.0_U0.00_gg0.60_ww0.50_T0.010_Mw800_Ch0.50-0.40_Nph25_err70.00E-4_dG8.00E-2_Ne6_Ns6_CI6_CASOrb1-12.eps, width = 0.38\textwidth} }
\caption{
(a) The spectral function for $\Omega=0.6$, 0.5, 0.4 and 0.36 at half filling. 
A dip around zero is the indication of superconducting gap. 
As $\Omega$ decreases, and the superconducting gap becomes more apparent. 
Below the optimal Boson energy ($\Omega=0.36$), the DOS keeps on 
decreasing because of the increasing $Z$ [Fig.~\ref{fig:Half_Gap_g0.6}(b)],
which is a consequence of the strong polaron effect.
(b) The spectral function for $\Omega=0.5$ at fillings of 1 (solid curve with shaded region) and 0.8. 
A dip around zero is seen for both cases.
}
\label{fig:SpectralFunction_g0.6}
 \end{figure}

Fig.~\ref{fig:Half_Gap_g0.6}(a) shows the (half) superconducting gap as a function of Boson energy at the half filling. 
The electron-boson coupling is fixed at $g=0.6$. We find that there exists an optimal Boson energy $\Omega_{opt}$
for the superconducting gap, which is around $\Omega_{opt} \sim 0.4$ for these parameters. 
The ME theory, although resulting in different numerical values, exhibits the same non-monotonous behavior, with the 
optimal Boson energy at $\Omega_{opt} \sim 0.15$. 
To analyze the origin of the non-monotonous behavior, Fig.~\ref{fig:Half_Gap_g0.6}(b) shows $Z(i \omega_0)$ and $\phi(i \omega_0)$ 
(whose ratio determines the gap amplitude) as a function of Boson energies. 
The ME theory results in the similar but milder $Z(i \omega_0)$  and $\phi(i \omega_0)$ behavior.
At very large  $\Omega$ ($\Omega \gg \Omega_{opt}$), both polaron and superconducting effects are weak 
($Z(i \omega_0) \sim 1$ and $\phi(i \omega_0) \sim 0$), because the Boson energy is too large and has 
little effects on the ground state property. 
When $\Omega$ decreases, both $Z(i \omega_0)$  and $\phi(i \omega_0)$ increase, but at different rates.
At larger $\Omega$ ($\Omega \gtrsim \Omega_{opt}$), $\phi(i \omega_0)$ grows faster, leading to increasing gap amplitudes.
At smaller $\Omega$ ($\Omega \lesssim \Omega_{opt}$), $Z(i \omega_0)$ grows faster, leading to decreasing gap amplitudes.
As $\Omega \rightarrow 0 $ (only obtained using ME theory), 
both  $Z(i \omega_0)$  and $\phi(i \omega_0)$ diverge, with the former being much faster.
From our discussion in Section II.D, the decreasing superconducting gap below $\Omega_{opt}$ 
is the consequence that the polaron effect starts to dominate over the superconductivity at small $\Omega$.  
Fig.~\ref{fig:Ordere_g0.6} provides the pairing amplitude $\Psi (\equiv \langle d_\uparrow d_\downarrow \rangle)$ 
as a function of the Boson energy $\Omega$, and the shape similar the superconducting gap is seen. 
The spectral functions at half filling for $g=0.6$ are provided in Fig.~\ref{fig:SpectralFunction_g0.6}(a). 
As the Boson energy decreases, the spectral function first develops a dip \cite{ZeroDOS}
around the Fermi energy, indicating the superconducting state, and then keeps on decreasing in value
due to the increasing polaron effect  [$Z$ in Fig.~\ref{fig:Half_Gap_g0.6}(b)]. 
We emphasize that both $\phi$ and $Z$ lead to a reduction in the spectral function near $E_F$,
and it is not easy to distinguish the polaron from the superconducting effect from the spectral function alone.
An explicit evaluation of $Z$ and $\phi$ to separate these two effects. 
To confirm that the dip around zero for $\Omega>\Omega_{opt}$ is indeed caused by the superconductivity, not by the error caused 
by including only five bath orbitals, Fig.~\ref{fig:SpectralFunction_g0.6}(b) 
presents the spectral function computed at band filling of 0.8. 
A dip around zero is also clearly observed.

\begin{figure}[htp]
\epsfig{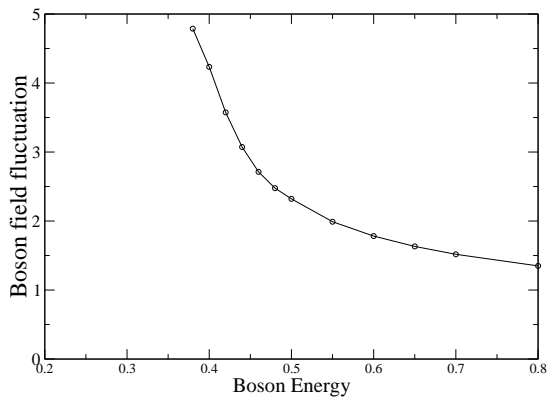} 
\caption{ $\sqrt{\langle (a+a^\dagger)^2 \rangle/\Omega}$, which 
measures the Boson field fluctuation, as a function of the Boson energy $\Omega$.
This value increases upon decreasing $\Omega$.
}
\label{fig:x^2_average}
 \end{figure}


\subsection{Limitations of solvers and comparison to other methods}


There are two dimensionless parameters that govern the validity of the ME theory -- one 
$\Omega/E_F$ and $\lambda = \frac{g^2}{\Omega t}$. The former is derived from the Midgal theory \cite{Migdal_1958, Grimvall, Allen_SC}
which gives the condition where the vertex correction can be neglected; the later is obtained 
from DMFT \cite{PhysRevB.58.14320, PhysRevB.65.224301}, and is proportional to $\left.\frac{\partial \Sigma(\omega)}{\partial \omega}\right|_{\omega=0}$
obtained from ME theory \cite{Allen_SC}. Both parameters have to be small for ME theory to work. 
Roughly, the small $\lambda$ guarantees the correct ground state, and the small $\Omega/E_F$ gives the correct excitations  \cite{PhysRevB.65.224301}.
Note that small $\lambda$ implies that ME theory is not valid for any given $g$ at small enough $\Omega$.
As DMFT becomes exact in the infinite-dimension limit, we use it as a reference to see when and how 
the ME theory breaks down. As expected, the gaps obtained from  ME theory and DMFT become closer
when $\lambda$ is small (larger $\Omega$ at a given $g$). When $\lambda$ is large, ME theory 
does not properly capture the polaron effect and thus gives the (wrong) superconducting ground state.
Without considering the superconducting solution, Ref.~\cite{PhysRevB.58.14320} determines the critical value $\lambda_c$
is of order one, above which the system becomes insulating. 
From Fig.~\ref{fig:Half_Gap_g0.6}(a) and  Fig.~\ref{fig:Ordere_g0.6}, we see that 
below $\Omega \sim 0.33$, the superconducting amplitude becomes negligible and the system is in the polaron state.
As the ME theory predicts the superconducting state for all $\Omega$, our calculation results in 
a $\lambda_c \approx \frac{0.6^2}{0.33 \times 1} \approx 1.1$, above which the ME theory gives the wrong ground state.
We have done the calculations for $g=0.3$, 0.4 and 0.5 (not shown), and the resulting $\lambda_c$ are all around 1.
Our DMFT calculations thus extend the criterion given in Refs.~\cite{PhysRevB.58.14320, PhysRevB.65.224301}
to the superconducting solution.

We also examine the harmonic Boson potential at small $\Omega$.
We first represent the Boson field as a simple harmonic oscillator, i.e.
$\Omega (a^\dagger a + 1/2) = \frac{p^2}{2m} + \frac{m \Omega^2 }{2} x^2$ (the convention $\hbar \equiv 1$ is used), with $x$ a distortion field. 
In this representation the distortion $x = \frac{1}{\sqrt{2 m \Omega} } (a^\dagger + a)$, 
and $\sqrt{ \langle x^2 \rangle} \sim \sqrt{\langle (a+a^\dagger)^2  \rangle/\Omega}$ characterizes the distortion fluctuation.
$\sqrt{\langle (a+a^\dagger)^2  \rangle/\Omega}$ as a function of $\Omega$ is given in Fig.~\ref{fig:x^2_average},
which clearly shows a divergent behavior at small $\Omega$. 
Note that the diverging behavior in $Z$ and $\phi$ [Fig.~\ref{fig:Half_Gap_g0.6}(b)], and the diverging behavior
of $\sqrt{\langle (a+a^\dagger)^2  \rangle/\Omega}$ [Fig.~\ref{fig:x^2_average}] happen at the same $\Omega$,
which signals the polaron insulating phase.
When $\sqrt{ \langle x^2 \rangle}$ is comparable to the lattice constant or the inter-electron distance,
the validity of the quadratic potential may not be sufficient.

We now compare our results to those in the literature. 
We first discuss the DMFT results using other impurity solvers, including the Hirsch-Fye \cite{PhysRevLett.56.2521} Quantum Monte Carlo (QMC)
\cite{PhysRevB.48.6302, PhysRevB.50.6939, PhysRevLett.75.2570}, the second order perturbation in phonon propagators \cite{PhysRevB.50.6939},
the semiclassical solver \cite{PhysRevB.54.5389, PhysRevB.79.205109}, the path integral \cite{PhysRevB.58.14320},  
the diagrammatic expansion \cite{PhysRevB.65.224301}, and the continuous-time QMC \cite{RevModPhys.83.349, PhysRevLett.113.266404}. 
The Hirsch-Fye QMC is formally exact, and works efficiently at high temperature \cite{PhysRevB.48.6302, PhysRevB.50.6939, PhysRevLett.75.2570}. 
The critical temperatures for charge density wave (CDW) 
and superconducting phases are determined by the divergence of the corresponding susceptibilities. 
The ED solver used here is for zero-temperature phases, and a direct comparison cannot be made. 
An investigation on CDW order at zero temperature is worthwhile.
We notice that in the large $\lambda$ limit, DMFT yields the polaron insulating solution, which 
can easily lead some CDW order as  the local occupation prefers either zero or two electrons.  
In this sense, the DMFT phase diagram is consistent with the results from the QMC calculation,
where the CDW phase happens at small $\Omega$ regime whereas superconductivity at large $\Omega$ regime \cite{PhysRevB.48.6302}. 
The semiclassical solver \cite{PhysRevB.54.5389, PhysRevB.79.205109} neglects the Boson dynamics, 
and captures only the polaron but not the superconducting physics.
The analysis based on path integral \cite{PhysRevB.58.14320} and diagrammatic expansion \cite{PhysRevB.65.224301}
identifies an important dimensionless parameter $\lambda = g^2/(\Omega t)$, and our results (discussed in the first paragraph in this Section) 
are fully consistent with these results. 
In the parameter regime where both $\Omega$ and $g$ are comparable to the bandwidth, 
the bipolaron effect becomes important \cite{PhysRevB.33.4526}, and several phases 
such as supersolid, CDW, and superconducting states, along with a quantum critical point are obtained \cite{PhysRevLett.113.266404}.
We do not explore the parameter regime, but it is the regime where the ED solver is applicable.
Finally we note that in the two-dimensional electronic system, a weak electron-phonon coupling can lead to the CDW order 
that can coexist with the superconducting state \cite{PhysRevB.40.197, PhysRevB.46.271, 0295-5075-85-5-57003}.
The CDW order exhibited in this case originates mainly from the nesting of the band, but has nothing to do with the specific form
the local interaction.

\subsection{Connection to the recent high-pressure experiment}
In 2015, Drozdov {\em et. al.} shows that applying a high pressure to on sulfur hydride (H$_2$S)
can enhance the superconducting $T_c$ up to 203K, and the isotope effect (replacing hydrogen by deuterium and tritium reduces $T_c$)
further confirms that it is the phonon-mediated conventional superconductor \cite{Nature525.73.2015}. 
One of their findings is that there is an optimal pressure for the superconductivity,
above which the superconducting $T_c$ starts to decrease. 
Our calculation suggests a simple explanation for this non-monotonous behavior as a function of pressure,
under the following three (reasonable) assumptions: (i) the Boson energy $\Omega$ considered in the Holstein model corresponds 
to the Debye frequency (multiplied by $\hbar$);  (ii) applying a pressure increases the 
Debye frequency; and (iii) the electron-phonon coupling does not change significantly (within 10$\%$) upon applying the pressure. 
The isotope effect, which lowers the Debye frequency via increasing the atomic masses, lowers the superconducting $T_c$. 
This well-known behavior corresponds to the Boson energy smaller than the optimal $\Omega_{opt}$. 
When the applied pressure is too large such that the Debye frequency passes its optimal value,
the superconducting $T_c$ again decreases. We emphasize that 
the optimal pressure can also be caused by other physics -- for example, the Coulomb repulsion
becomes stronger upon increasing the pressure, and leads to a reduction of superconducting $T_c$. 
Our calculation cannot tell which one is the main mechanism. However, it does imply
that an optimal pressure exists even without invoking the Coulomb repulsion. 
To quantitatively see how superconductivity is affected by a Hubbard $U$ is worthy of further investigations.

\section{Conclusion}

In this work we examine the superconducting solution in the Holstein model with semicircular density of states, 
using both the Migdal-Eliashberg theory and Dynamical Mean Field theory. 
The impurity model associated with  DMFT is solved using the exact diagonalization.
Although different in numerical values, both methods imply that for a given 
electron-Boson coupling there exists an optimal Boson energy for superconductivity.
By analyzing the Green's function, this non-monotonous behavior originates from 
the interplay between superconducting and polaron effects. 
At large $\Omega$, the polaron effect is small so the superconducting gap increases upon lowering $\Omega$.
Below certain $\Omega$, the polaron effect starts to dominate and therefore reduces the superconductivity by making the system
less metallic (reducing the DOS around the Fermi energy). 
In terms of  many-body solvers, our DMFT results explicitly confirm that in the small $\Omega$
limit, the ME theory breaks down by getting the wrong ground state.
This result was already obtained in the calculations without breaking symmetries \cite{PhysRevB.58.14320, PhysRevB.65.224301},
and here we extend this statement to the superconducting solution.
Our calculation provides a simple explanation on the recent experiment on sulfur hydride, 
where a optimal pressure for the superconductivity was observed \cite{Nature525.73.2015}. 
Searching Boson degrees of freedom (other than the phonons) to mediate the electron-electron attraction 
can be a promising approach to enhance the superconducting temperature.

\subsection*{Acknowledgement}
We thank Qi Chen and Prabhakar Bandaru for helpful discussions, and Andrew Millis
for very insightful comments.
 

\bibliography{SC_meta}

\begin{thebibliography}{64}
\bibitem[{Zer()}]{ZeroDOS}
\bibinfo{note}{Due to the small number of bath orbitals, the zero DOS cannot be
  observed in the calculation.}
\end{thebibliography}

\end{document}